\documentclass[conference]{IEEEtran}
\IEEEoverridecommandlockouts
\usepackage{cite}
\usepackage{amsmath,amssymb,amsfonts}
\usepackage{algorithmic}
\usepackage{graphicx}
\usepackage{textcomp}
\usepackage{xcolor}
\usepackage{todonotes}
\usepackage{framed}
\usepackage{float}
\usepackage[export]{adjustbox}
\usepackage{caption}
\usepackage{subcaption}

\usepackage[colorlinks=true, linkcolor=black,urlcolor=black, citecolor=black]{hyperref}
\def\BibTeX{{\rm B\kern-.05em{\sc i\kern-.025em b}\kern-.08em
    T\kern-.1667em\lower.7ex\hbox{E}\kern-.125emX}}
\begin{document}

\title{A VR Serious Game to Increase Empathy towards Students with Phonological Dyslexia \\

}

\newcommand{\newlineauthors}{%
  \end{@IEEEauthorhalign}\hfill\mbox{}\par
  \mbox{}\hfill\begin{@IEEEauthorhalign}
}
\author{
\IEEEauthorblockN{José M. Alcalde-Llergo}
\IEEEauthorblockA{
\textit{Dept. of Economics, Engineering,} \\
\textit{Society and Business Organization (DEIM)}\\
\textit{University of Tuscia }\\
Viterbo, Italy \\
jose.alcalde@unitus.it}
\and
\IEEEauthorblockN{Enrique Yeguas-Bolívar}
\IEEEauthorblockA{\textit {Computing and Numerical Analysis} \\
\textit{University of Córdoba}\\
Córdoba, Spain \\
eyeguas@uco.es}
\and
\IEEEauthorblockN{Pilar Aparicio-Martínez}
\IEEEauthorblockA{\textit{Nursing, Physiotherapy and Pharmacology} \\
\textit{University of Córdoba}\\
Córdoba, Spain \\
n32apmap@uco.es}
\and
\IEEEauthorblockN{Andrea Zingoni}
\IEEEauthorblockA{\textit{Dept. of Economics, Engineering,} \\
\textit{Society and Business Organization (DEIM)}\\
\textit{University of Tuscia}\\
Viterbo, Italy \\
andrea.zingoni@unitus.it}

\and
\IEEEauthorblockN{Juri Taborri}
\IEEEauthorblockA{
\textit{Dept. of Economics, Engineering,} \\
\textit{Society and Business Organization (DEIM)}\\
\textit{University of Tuscia }\\
Viterbo, Italy \\
juri.taborri@unitus.it}

\and
\IEEEauthorblockN{Sara Pinzi}
\IEEEauthorblockA{\textit {Dept. Physical Chemistry} \\
\textit{and Applied Thermodynamics}\\
\textit{University of Córdoba}\\
Córdoba, Spain \\
qf1pinps@uco.es}
}

\maketitle

\begin{abstract}
Dyslexia is a neurodevelopmental disorder that is estimated to affect about 5-10\% of the population. In particular, phonological dyslexia causes problems in connecting the sounds of words with their written forms. This results in difficulties such as slow reading speed, inaccurate reading, and difficulty decoding unfamiliar words. Moreover, dyslexia can also be a challenging and frustrating experience for students as they may feel misunderstood or stigmatized by their peers or educators. For these reasons, the use of compensatory tools and strategies is of crucial importance for dyslexic students to have the same opportunities as non-dyslexic ones. However, generally, people underestimate the problem and are not aware of the importance of support methodologies. In the light of this, the main purpose of this paper is to propose a virtual reality (VR) serious game through which teachers, students and, in general, non-dyslexic people could understand which are some of the issues of student with dyslexia and the fundamental utility of offering support to them. In the game, players must create a potion by following a recipe written in an alphabet that is specifically designed to replicate the reading difficulties experienced by individuals with dyslexia. The task must be solved first without any help and then by receiving supporting tools and strategies with the idea that the player can put himself in the place of the dyslexic person and understand the real need for support methodologies.

\end{abstract}

\begin{IEEEkeywords}
Virtual reality, Simulation, Inclusion, Dyslexia, Empathy, Serious game
\end{IEEEkeywords}

\section{Introduction}
Dyslexia is a specific learning disorder that causes significant difficulties in learning skills related to reading \cite{what_SLD_Dyslexia}. It is estimated to affect approximately 5-10\% of the population \cite{dyslexia_adults_review}, which is equivalent to about 700 million people worldwide.


More specifically, phonological dyslexia is an impairment of reading novel words (non-words) with otherwise good performance in reading familiar words~\cite{phon_dys}. Individuals with phonological dyslexia often experience problems to connect the sounds of words to their written forms, which can hinder their ability to recognize and recall new words. Phonological dyslexia is thought to be caused by differences in the way the brain processes language, particularly in the regions of the brain that are responsible for phonological processing~\cite{phon_dys_causes}. This can result in reading difficulties, such as slow reading speed, inaccurate reading, and difficulty with decoding unfamiliar words \cite{phon_dyslex_dysg}. However, with appropriate support and interventions individuals with phonological dyslexia can compensate these problems and can achieve success in their academic and professional lives. One of this support is represented by the inclusion of dyslexic students by their classmates and teachers.

Thus, empathy towards dyslexic students is of fundamental importance and stimulating it is one of the main objectives of this work. Indeed, dyslexia can be a challenging and frustrating experience for students as they may feel misunderstood or stigmatized by their peers or educators. Demonstrating empathy towards students with dyslexia can help to create a supportive and inclusive learning environment, where they feel understood, validated, and respected \cite{dys_frustration}. This can help them to feel more confident and motivated to learn, and ultimately lead to greater success in their academic and professional lives.

To achieve this, we propose a VR serious game to promote understanding of dyslexia and make people more conscious of the problem that it can create. This should foster the creation of a supportive environment that enables study success for dyslexic students.

The present work emerges as part of the VRAIlexia project~\cite{VRAILEXIA}, an initiative designed to increase awareness and provide support for university students with dyslexia. The main objective of the project is to mitigate the challenges arising from dyslexia among students in higher education, with the goal of reducing the incidence of university dropouts and facilitating access to degree programs for individuals with dyslexia, a severe problem addressed in \cite{juri}.

\section{Related works}

\subsection{Virtual reality for inclusion}
Virtual reality (VR) technology has the potential to create immersive and interactive environments that can simulate real-world experiences. This technology can be applied to all kinds of fields, including education, healthcare, entertainment, and social sciences. In particular, VR has shown promising results in promoting social inclusion as can be seen in works such as the Includiamoci project~\cite{VR_social_inclusion}. It is a social inclusion initiative that utilizes VR and spatial augmented reality technologies to create safe spaces for collaborative activities based on art therapy techniques and new technologies. The project aims to recognize and enhance individual differences while promoting teamwork and continuous confrontation between participants, educators, and experts in cultural heritage and technologies.

VR has also been applied to achieve the inclusion of disabled people into the labor market~\cite{laboral_inclusion}. In this case, authors use immersive environments to help people with disabilities to increase employment opportunities. The system aims to facilitate the integration of individuals with disabilities into the labor market by simulating work activities and enabling the development of skills in a pleasurable and active way. The software also serves as an alternative communication system, promoting social integration and cognitive development.

In other fields, VR has been used for inclusion by proposing activities in virtual environments that help working on the difficulties faced by some social groups. An example of this can be seen in a case study where authors investigated the impact of a ``Virtual Reality Social Cognition Training'' to enhance social skills in children with Autism Spectrum Disorder (ASD)~\cite{vr_autismo}. During the study, the performance of 30 children with ASD was measured in different domains by putting them in different situations such as doing a team project in the classroom or ordering food in the school cafeteria. The study findings indicated that utilizing a virtual reality platform is a promising treatment approach to understanding better social impairments frequently observed in individuals with ASD. Specifically, enhancements in emotion recognition, social attribution, and executive function of analogical reasoning where shown.

\subsection{Virtual reality for empathy}
Another application where VR can exploit its potential is in facilitating empathy towards a social group. There are studies that prove the beneficts of VR in this task, such as the performed in~\cite{vr_empathy}. In this study participants where chosen randomly to view a documentary featuring a young girl living in a refugee camp in a VR format. They found that VR can lead to greater experience of empathy.

Also a VR application as a means to empathize a case study focused in wheelchair users was developed in~\cite{wheelchairs}. The objective of this study was to analyze the effects of a simulation aimed at replicating the challenges that a student who uses a wheelchair would face when performing various tasks in their daily lives. The results showed that the simulation experience changed attitudes towards persons with disabilities in the real world and made them advocates for change. The study adds to existing research on the importance of VR simulations in improving empathy towards others.

\subsection{Virtual reality for dyslexia}
The use of VR to address some of the problems caused by dyslexia in children has also been a frequent topic of research in recent years. In addition, studies such as~\cite{pedroli2017psychometric} have shown that the use of VR can improve the memory and skills of people with dyslexia. In this case, users were subjected to different tests in a virtual classroom where they had to perform according to what appeared on the blackboard. The results did not show clear improvements in their reading skills, but they showed a clear improvement in their attention, which could lead, in the longer term, to improvements in other problems caused by dyslexia, as reducing time in reading low frequency long words.

A clear case study where dyslexia is dealt using VR can be seen in the European project FORDYS-VAR~\cite{fordysvar}, whose main objective is to offer a way to facilitate the learning of people with dyslexia through technology, more specifically using VR and augmented reality (AR). This project is especially focused on helping dyslexic children between 10 and 16 years old. Among its contributions, authors develop a support software for children with dyslexia \cite{fordysvar_tool} with which they can work to alleviate several of the learning disabilities caused by this disorder in a more entertaining way, through a virtual reality video game on the Oculus Quest platform.

It is worth noting that most of the studies on VR applied to dyslexia do not consider people at the higher education stage. Conversely, VRAIlexia project aims to help these students through VR and artificial intelligence technologies~\cite{VRAILEXIA}. During the project different VR applications were designed and developed, including an application for the realization of psychometric tests for students with dyslexia~\cite{metrox2022}, as well as the serious game that we present during the current document.


\section{Methodology}
The methodology proposed in this study concerns making players, such as teachers or peers of dyslexic students, feel inside a virtual world where they have to complete a task. The task requires participants to read a recipe book and correctly adding ingredients in the order and quantity specified. However, to simulate the reading difficulties experienced by people with dyslexia, the text in the book and ingredient labels are presented in the Britton’s Dyslexia font \cite{britton}, an alphabet that eliminates certain parts of letters, making reading more challenging. Figure \ref{fig:britton} shows how the word ``Dyslexia'' is written using this alphabet.

\begin{figure}[h]
    \centering
    \includegraphics[width=0.25\textwidth]{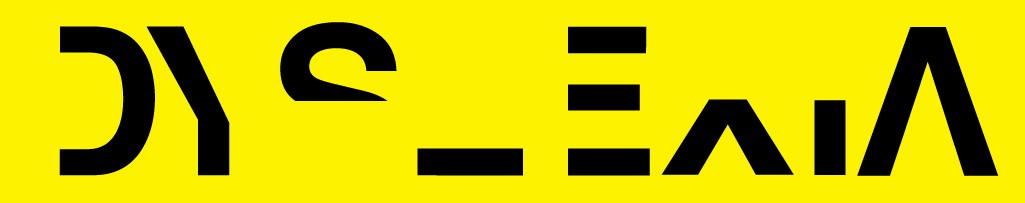}
    \caption{`The word `DYSLEXIA'' using Britton's font \cite{britton}.}
    \label{fig:britton}
\end{figure}

\subsection{Application flow}

Figure \ref{fig:flow_diag} shows the application flow of ``In the shoes of dyslexic students (Potion)''.

\begin{figure}[h]
    \centering
    \includegraphics[width=0.47\textwidth]{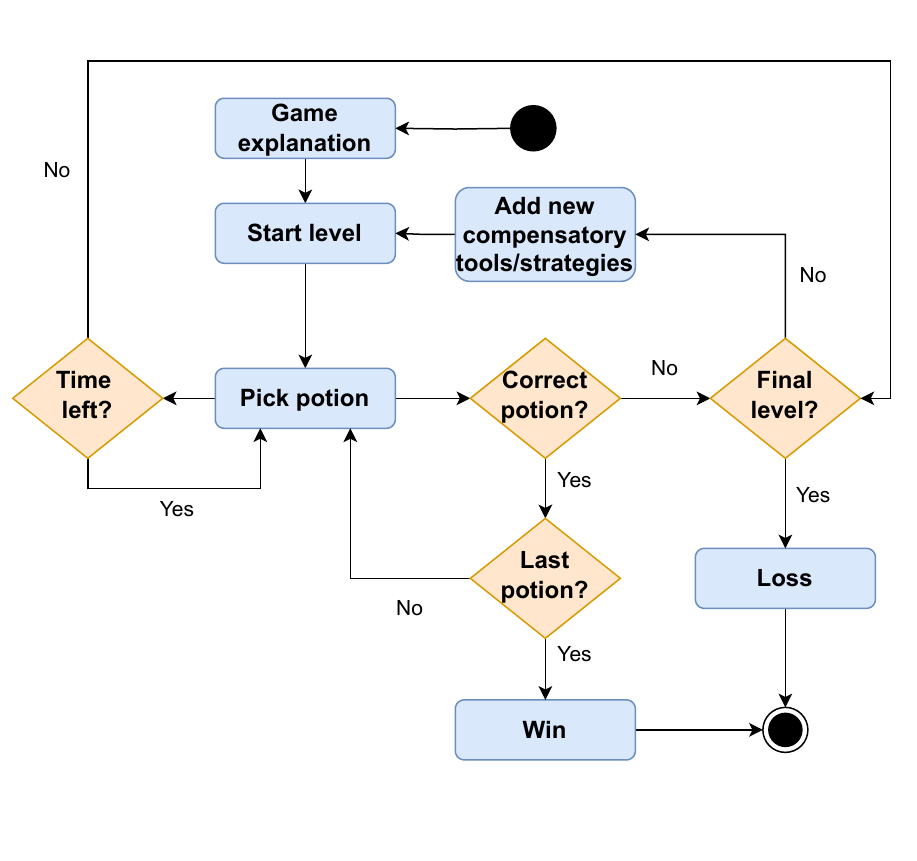}
    \caption{Flow diagram.}
    \label{fig:flow_diag}
\end{figure}

Once players enter the game they find themselves inside a kind of ``magic castle'', in front of a closed door. The reason behind using a fantasy design for the virtual rooms was two-fold: its widespread popularity and its ability to enhance the game's allure while also providing a purpose for the task. When they go through the door they meet their potions teacher, an avatar that will tell them that they must prepare a potion to save their friend, called Sam. She explains that the potion can be created by adding different ingredients strictly in the correct order into a pot, according to the recipe given in a book. Moreover, a time limit of 3 minutes must be respected. Ingredients are distributed throughout the room on different shelves. 

After that, players initiate their first attempt to save Sam. As they fail these attempts, they will be given more time, 5 or 10 minutes, and new compensatory tools, like shorter words and an audio guide, to try saving their friend again. A level failure occurs when time runs out or when a wrong ingredient is put into the pot or, also, a correct ingredient is dropped in an incorrect order. The game will end once the players fail all three attempts given or succeed in making the potion correctly during one of them. 

\subsection{Virtual rooms}

Within the virtual castle, participants will be granted admittance to three distinct rooms. The initial of these enclosures is the starting point of our exploration, where players can familiarize themselves with the operational features of the controller buttons. In addition, they will find a brief explanation about the effects of phonological dyslexia as shown in Figure \ref{subfig:start_room}. Second room is Sam's room. It is an empty room where players can see their friend and his current state as shown in Figure \ref{subfig:sam_room}. Finally, the third room is the one where the game takes place. It is a potions laboratory where players can find their teacher and all the materials needed to brew a potion to help Sam. A general view of the room is shown in Figure \ref{subfig:potions_room}.

\begin{figure*}
        \centering
        \begin{subfigure}[b]{0.302\textwidth}
             \centering
             \includegraphics[width=\textwidth]{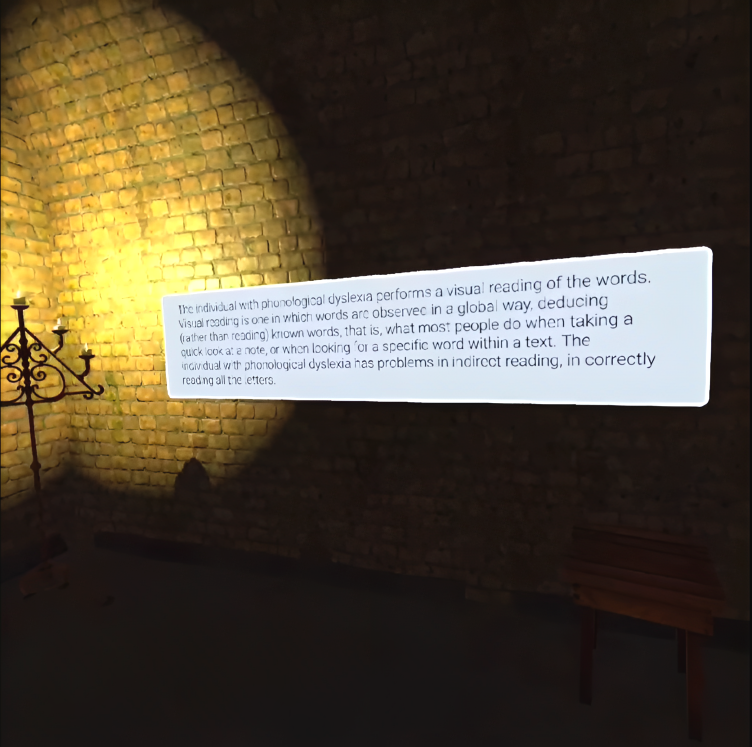}
             \caption{Starting room.}
             \label{subfig:start_room}
        \end{subfigure}
        \begin{subfigure}[b]{0.3\textwidth}
             \centering
             \includegraphics[width=\textwidth]{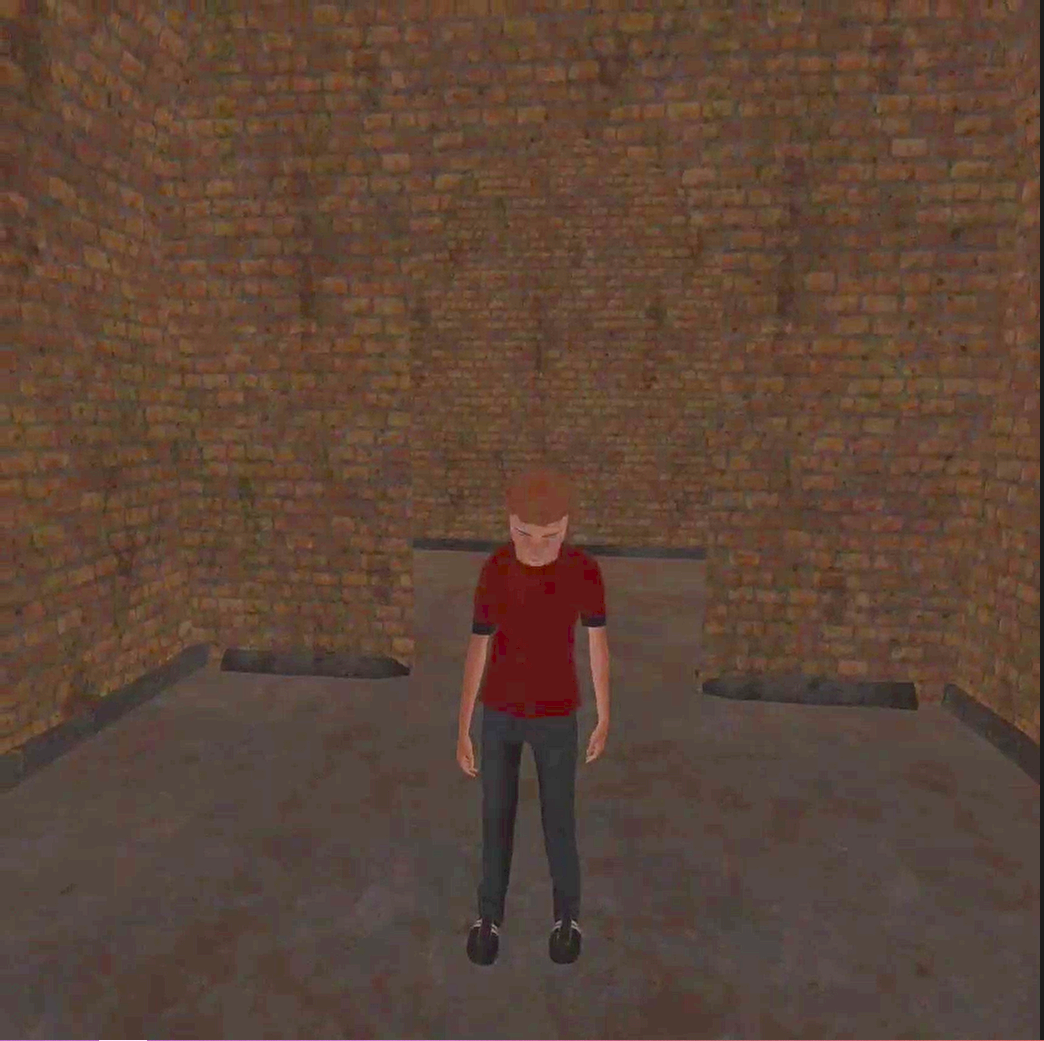}
             \caption{Sam's room.}
             \label{subfig:sam_room}
        \end{subfigure}
        \begin{subfigure}[b]{0.3\textwidth}
             \centering
             \includegraphics[width=\textwidth]{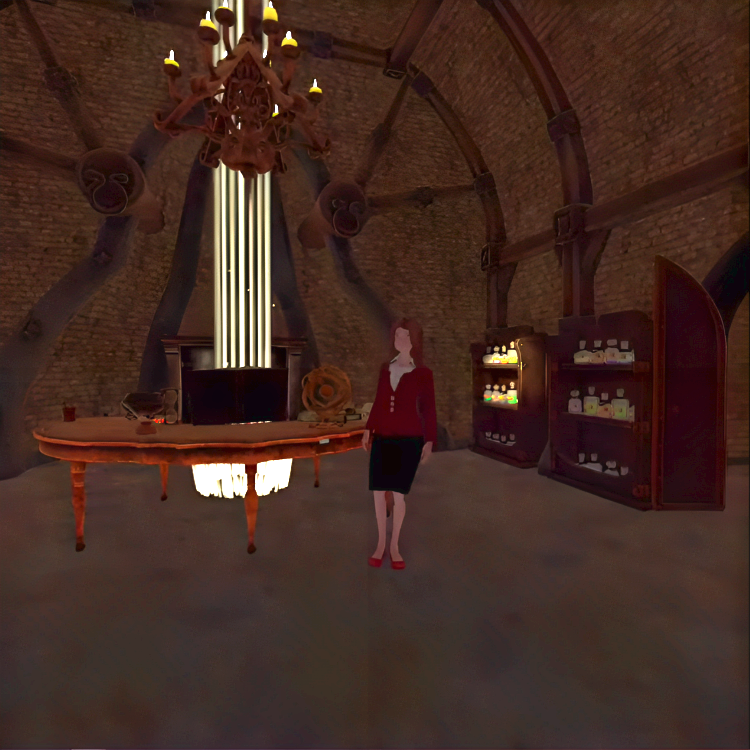}
             \caption{Potions laboratory.}
             \label{subfig:potions_room}
        \end{subfigure}
        \caption{Chambers from the virtual castle: initial room (a), Sam's Room (b) and potions laboratory (c).}
        \label{fig:1}
    \end{figure*}

\subsection{Main characters and items}

Throughout the VR experience, participants are required to interact with a diverse range of items and characters that have been specifically designed to guide and enhance their level of immersion within the simulated environment. A list of these characters and objects is exposed below, and their representation in the virtual world are shown in Figure \ref{fig:main_items}.

\begin{itemize}
    \item \textbf{Teacher}: she has the role of potions teacher in the virtual world. She explains to the players the rules of the game belonging to the different levels and what they have to do to create the correct potion and win the game. In addition, she takes on the role of a strict teacher who yell at her students when they fail in their tasks. In this way the player put himself in the place of dyslexic students who are not valued for their effort by their teachers.
    \item \textbf{Sam}: he represents the player's friend in the game. The player must prepare a potion correctly and within a limited time frame in order to save him. Player's results will be directly reflected in Sam. If the correct potion is created Sam will be happy and celebrate, but if it fails during the different levels players will see him suffer.
    \item \textbf{Ingredients and shelves}: around the potions laboratory the player finds several shelves with different ingredients to make potions. These ingredients are carefully arranged in a particular order on each shelf, and stored within flasks of varying shapes and colors, each labeled with the corresponding ingredient's name. Notably, the labeling is composed utilizing the Britton's font, trying to replicate the reading difficulties experienced by a dyslexic person.
    \item \textbf{Table}: in the center of the potion laboratory lies an irregularly-shaped table designed to promote different modes of locomotion in VR (teleportation-based and controller-based). Over the table there are different key objects for the game, including an hourglass, recipe book, and pot, which players may interact with as part of their gameplay experience. Furthermore, the table serves as a practical space for players to place some of the potion ingredients. 
    \item \textbf{Hourglass}: is the item used to start the game. Next to it there is a digital clock with the time available to complete the level. Upon starting the game, the allotted time begins to decrement, with the level concluding once the timer reaches zero.
    \item \textbf{Pot}: container in which the potion is to be brewed. Players must pour the ingredients into the pot in the correct order to pass the game. The pot will release a heart if the poured ingredient is correct and a purple smoke if it is not, serving as feedback to guide the player towards successful completion of the game.
    \item \textbf{Recipe book}: it is a big book containing the recipe for the potion to be brewed by the player. It denotes the specific type and quantity of each ingredient required, and specifies the correct order in which they must be added to the container in order to ensure the successful completion of the potion. Consistent with the font utilized for ingredient labeling, the recipe book also employs the Britton's font.
    \item \textbf{Beacon}: a light beam that guides the player throughout the virtual environment. It indicates specific points within the potions laboratory where the player must be situated to initiate teacher conversations or start the next level.
\end{itemize}

\begin{figure*}[h]
        \centering
        \begin{subfigure}[b]{0.27\textwidth}
             \centering
             \includegraphics[width=\textwidth]{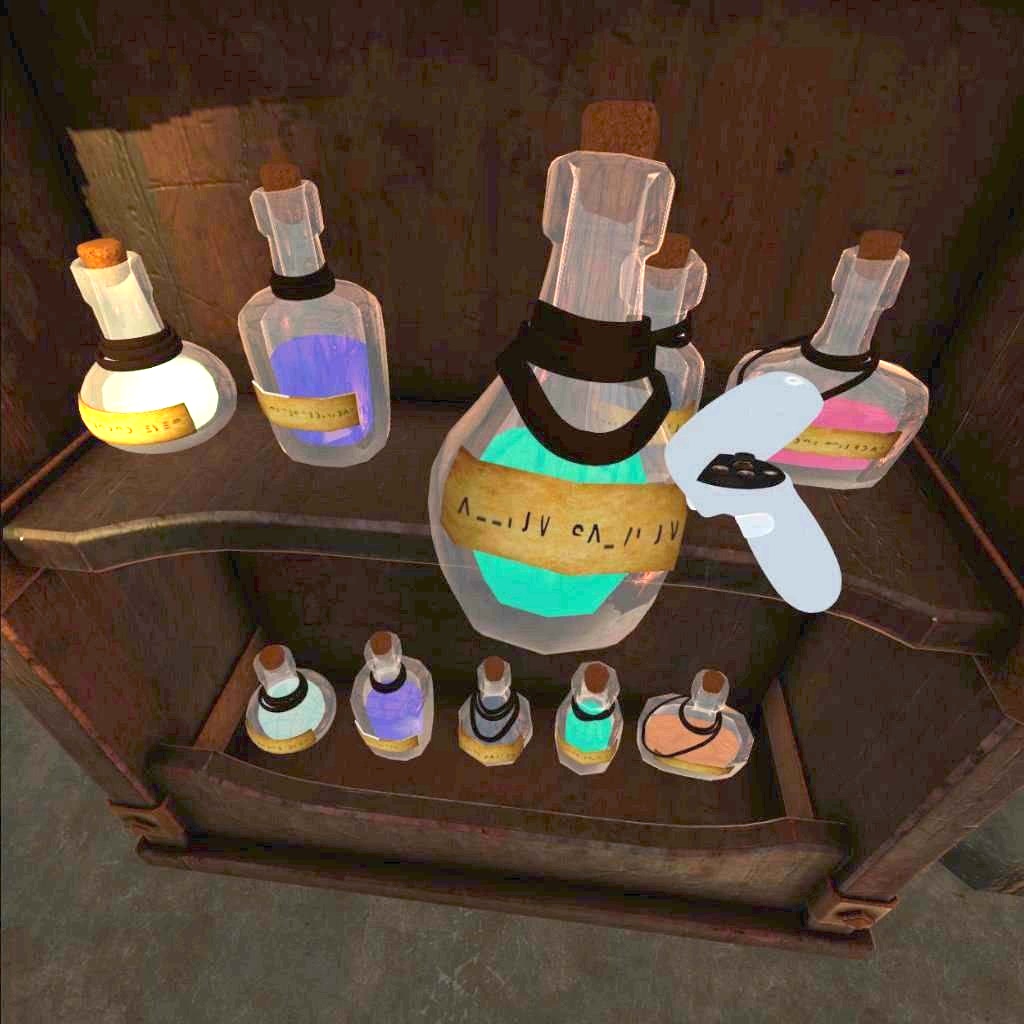}
             \caption{Ingredients.}
             
             \label{subfig:shelve}
        \end{subfigure}
        \begin{subfigure}[b]{0.27\textwidth}
             \centering
             \includegraphics[width=\textwidth]{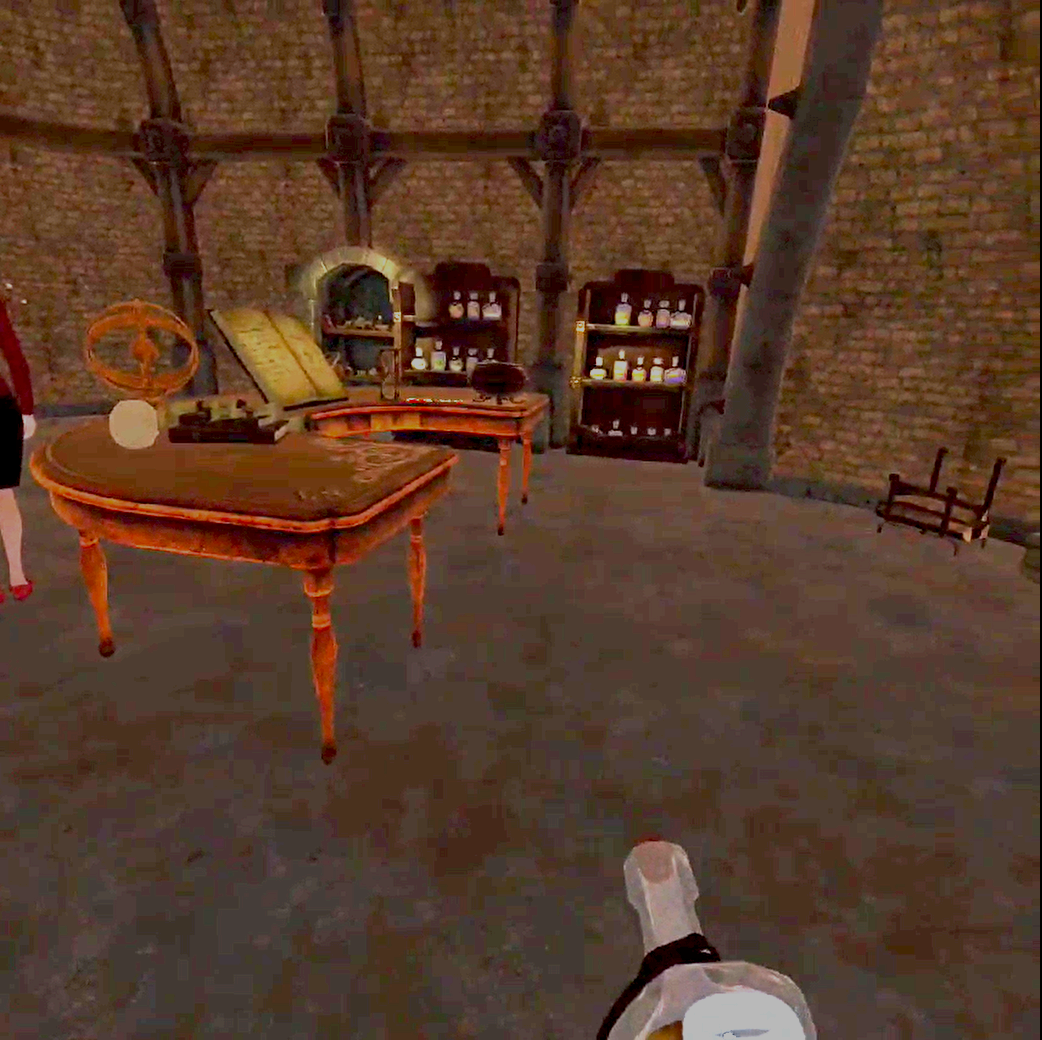}
             \caption{Irregular table.}
             \label{subfig:table}
        \end{subfigure}
        \begin{subfigure}[b]{0.27\textwidth}
             \centering
             \includegraphics[width=\textwidth]{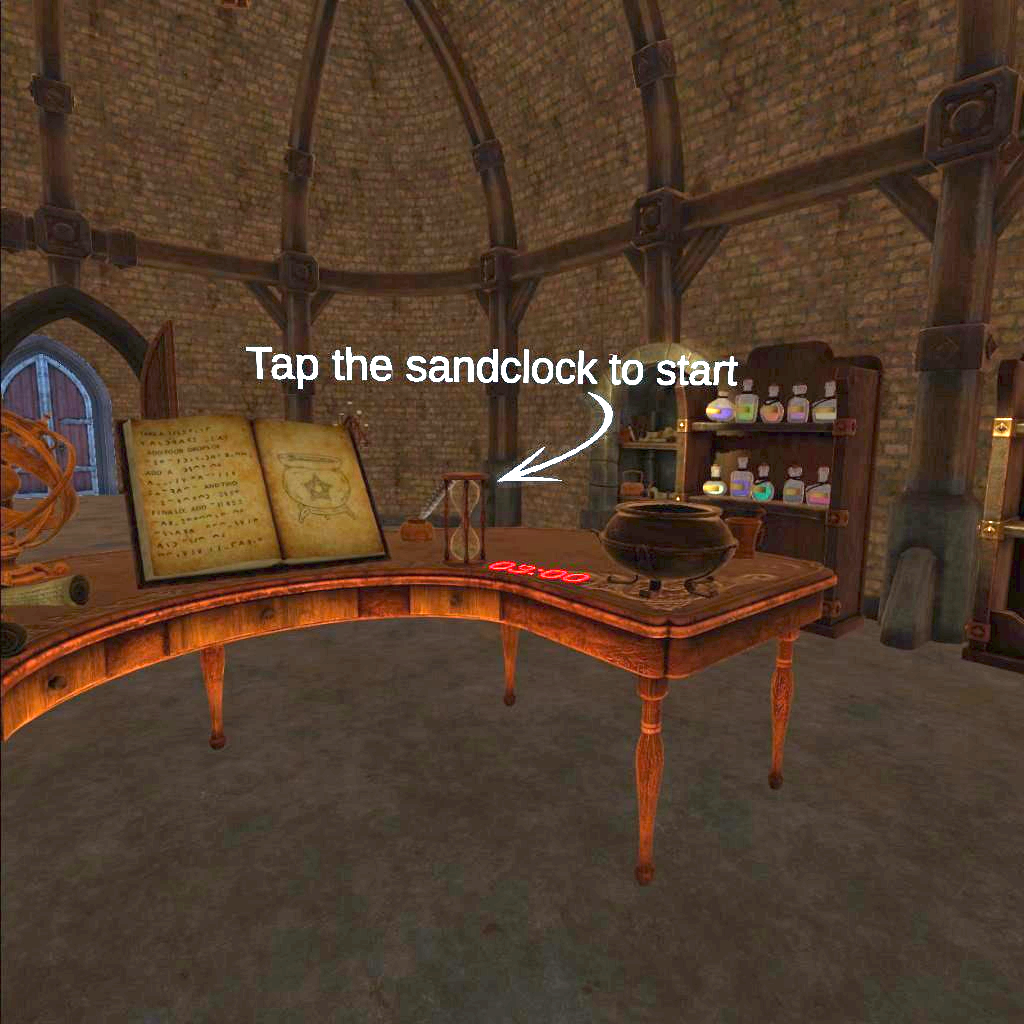}
             \caption{Hourglass.}
             \label{subfig:hourglass}
        \end{subfigure}
        \begin{subfigure}[b]{0.2701\textwidth}
             \centering
             \includegraphics[width=\textwidth]{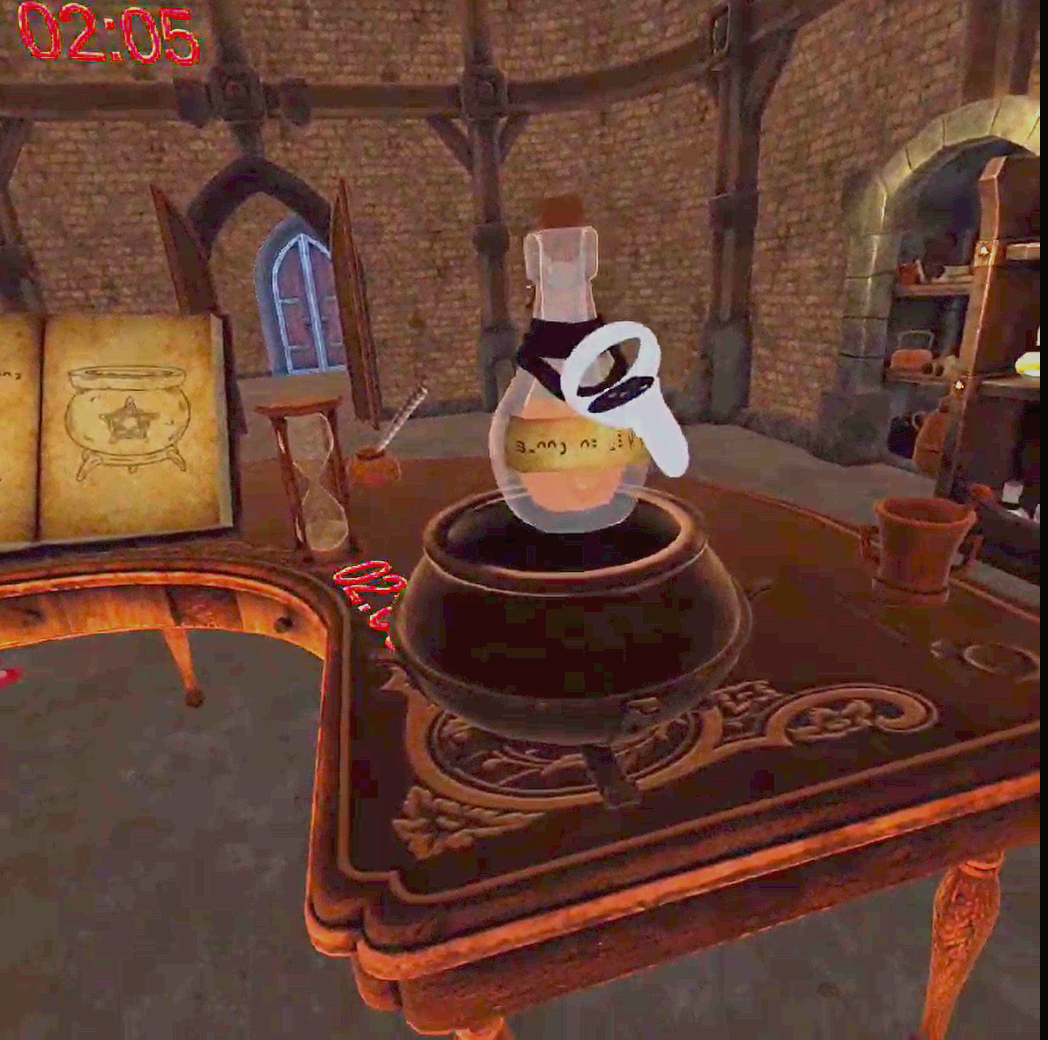}
             \caption{Pot.}
             \label{subfig:pot}
        \end{subfigure}
        \begin{subfigure}[b]{0.27\textwidth}
             \centering
             \includegraphics[width=\textwidth]{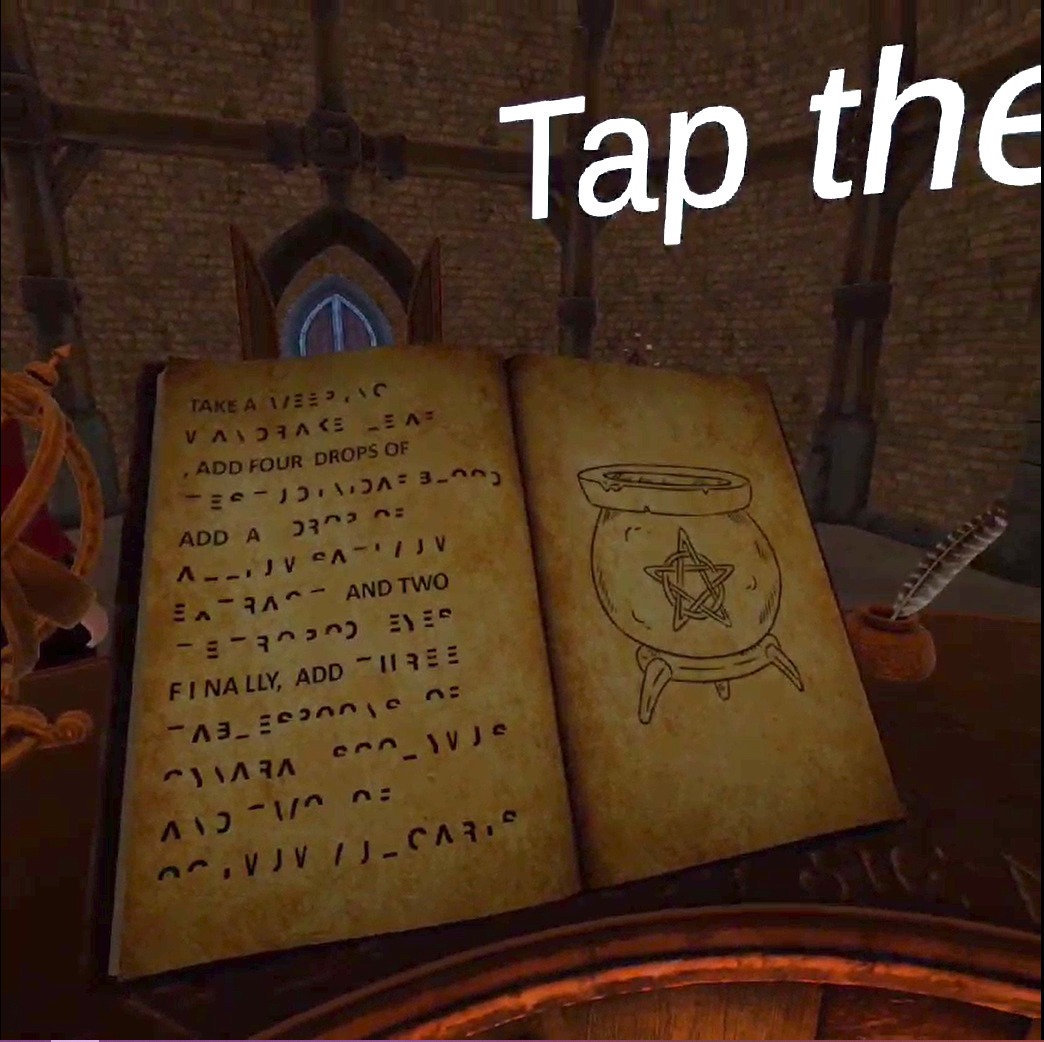}
             \caption{Recipe book.}
             \label{subfig:book}
        \end{subfigure}
        \begin{subfigure}[b]{0.27\textwidth}
             \centering
             \includegraphics[width=\textwidth]{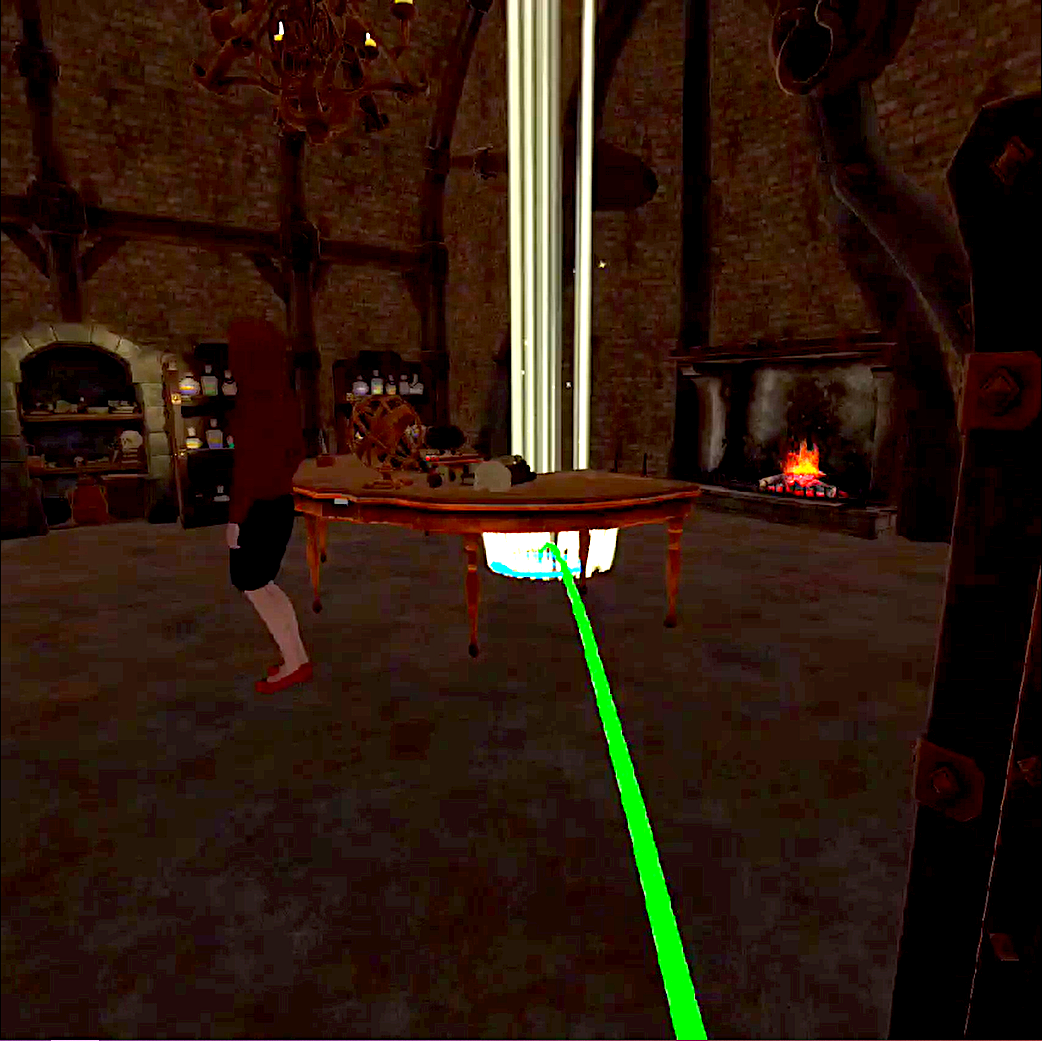}
             \caption{Beacon.}
             \label{subfig:beacon}
        \end{subfigure}
        \caption{Main items.}
        \label{fig:main_items}
    \end{figure*}

\section{Preliminary Results and Future Works}
The game was tested on 32 non-dyslexic individuals, who were not able, in general, to solve the task proposed, reporting feeling frustrated and anxious due to the difficulties in reading the instructions to complete the task. This feeling tended to disappear completely when more time was given and, above all, when compensatory tools were provided. In this way, consciousness about issues and needs of dyslexic students arose. People's opinions on the experience were collected through a survey with questions to be answered on a Likert scale ranging from 1 to 5. Figure \ref{fig:surv} shows two of these questions, the first referring to the difficulty of performing the proposed task and the second to the growth of their empathy towards people with dyslexia. The VR experience presented in this work can be thus considered a good way to increase empathy towards people with dyslexia.

\begin{figure}[h]
        \centering
        \begin{subfigure}[b]{0.385\textwidth}
             \centering
             \includegraphics[width=\textwidth]{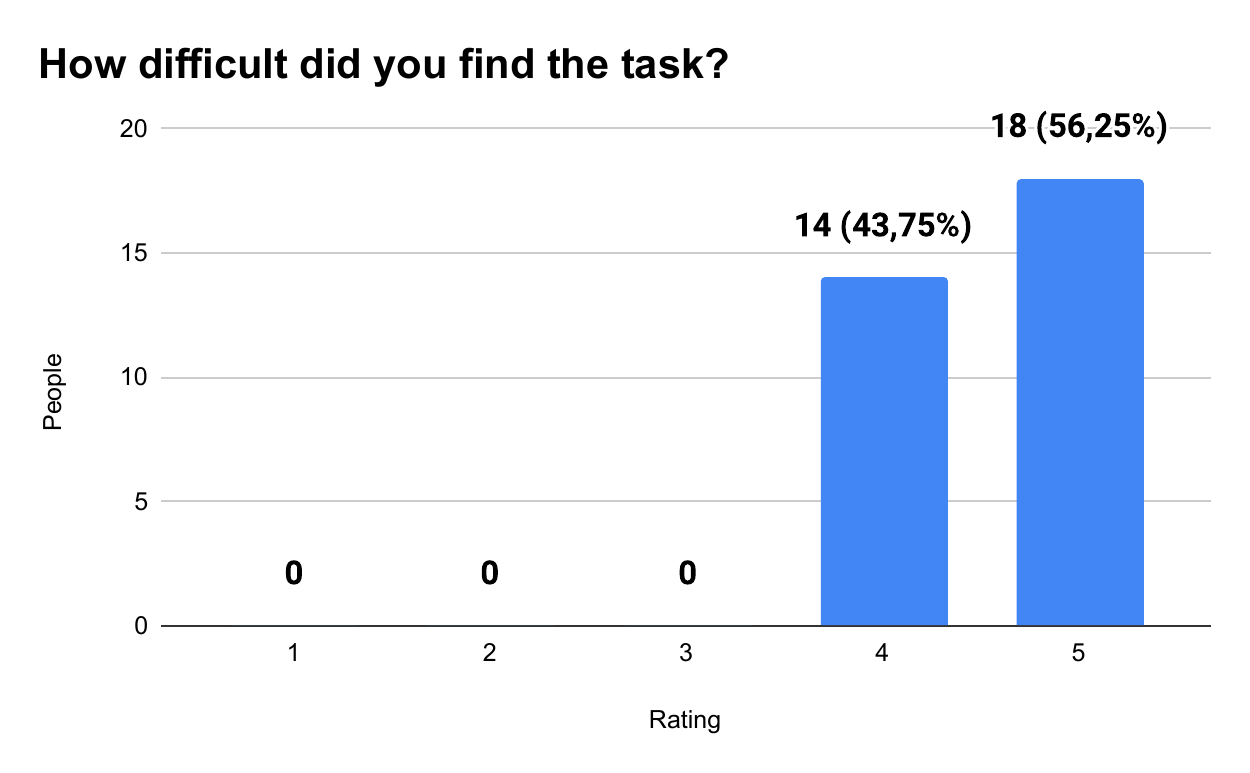}
             
             \label{subfig:shelve}
        \end{subfigure}
        \begin{subfigure}[b]{0.385\textwidth}
             \centering
             \includegraphics[width=\textwidth]{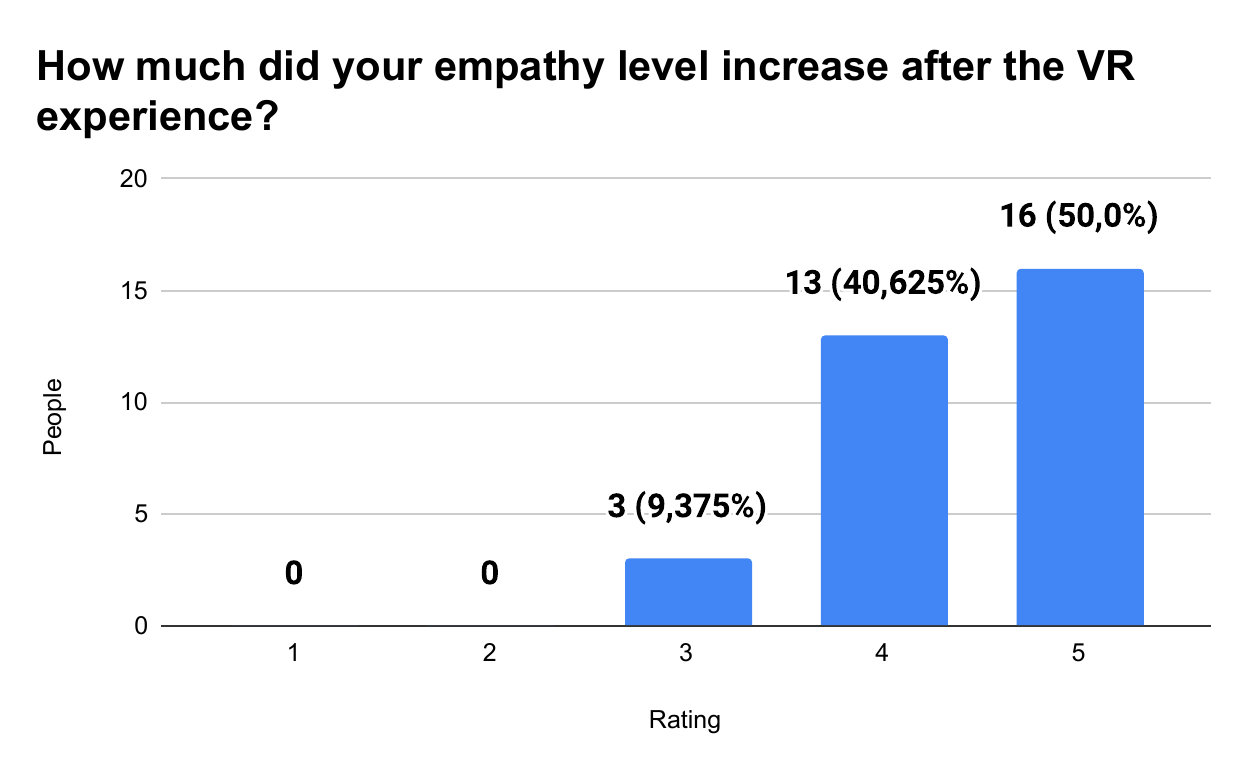}
        \end{subfigure}
        \caption{Survey results.}
        \label{fig:surv}
    \end{figure}

Next step involves collecting data from a larger sample size to validate the effectiveness of the developed serious game. Specifically, an appropriate survey will be developed to measure the quality of the VR experience and the increase in empathy towards people with phonological dyslexia after they complete the game. In addition, collected data will be used to study the different users empathy profiles through the application of various artificial intelligence techniques.

Finally, we are also designing other serious games to show empathy not only for people with phonological dyslexia, but also for those with other dyslexia types. In this way we have the objective that anyone can put themselves in the place of a dyslexic person, and thus understand the real need for the different methodologies of support required by all dyslexic students, including those in higher and university education.

\section{Acknowledgements}

José Manuel Alcalde Llergo is a PhD student enrolled in the National PhD in Artificial Intelligence, XXXVIII cycle, course on Health and life sciences, organized by Università Campus Bio-Medico di Roma. These results are framed in VRAIlexia project funded by the Erasmus+ Programme2014-2020 – Key Action 2: Strategic Partnership Projects. AGREEMENT n. 2020-1-IT02-KA203-080006.

\bibliographystyle{ieeetr}
\bibliography{bibliography}\clearpage

\end{document}